\documentclass[10pt,major,twocolumn]{IEEEtran}

\usepackage{pifont,epsfig,cite,psfrag,subfigure,enumerate,stfloats,balance}
\usepackage{graphicx}
\usepackage{subfigure}
\usepackage{subfig}
\usepackage{amsmath}
\usepackage{algorithm}
\usepackage{algorithmic}
\usepackage{cite}
\usepackage{amsthm}
\usepackage{amssymb}
\usepackage{color}
\usepackage{multicol}
\usepackage{bm}
\usepackage{tabularx}
\usepackage{booktabs}
\usepackage{epstopdf}
\usepackage{amsfonts}
\usepackage{setspace}
\usepackage{mathptmx}

\theoremstyle{definition}
\theoremstyle{remark}

\newtheorem{definition}{Definition}
\newtheorem{lemma}{Lemma}

\begin{document}
	\captionsetup[figure]{labelfont={},name={Fig.},labelsep=period}
	
\title{

\huge  Enhancing Mega-Satellite Networks with Generative Semantic Communication: A Networking Perspective

{\footnotesize 
}	
}

 \author{Binquan~Guo, Wanting~Yang, Zehui~Xiong,~\IEEEmembership{Senior Member, IEEE}, Zhou~Zhang,~\IEEEmembership{Member, IEEE}, Baosheng~Li, Zhu Han,~\IEEEmembership{Fellow,~IEEE}, Rahim Tafazolli,~\IEEEmembership{Fellow,~IEEE}, and Tony Q. S. Quek,~\IEEEmembership{Fellow,~IEEE}


\thanks{Binquan~Guo is the School of Telecommunications Engineering, Xidian University, Xi’an 710126, China, and Tianjin Artificial Intelligence Innovation Center (TAIIC), China, and also with  Singapore University of Technology and Design, Singapore 487372  (e-mail: bqguo@stu.xidian.edu.cn). }

\thanks{Wanting~Yang is with Singapore University of Technology and Design, Singapore 487372  (e-mail: wanting\_yang@sutd.edu.sg).}

\thanks{Zehui~Xiong  is with the School of Electronics, Electrical Engineering and Computer Science, Queen's University Belfast, BT9 5BN Belfast, U.K. (e-mail: z.xiong@qub.ac.uk).}

\thanks{Zhou~Zhang  is with College of Computer Science and Electronic Engineering, Hunan University, Changsha, P. R. China (e-mail: zt.sy1986@163.com).}

\thanks{Baosheng~Li  is with the School of Mathematics and Statistics, Xidian University, Xi' an, China (e-mail: bs.li@stu.xidian.edu.cn)}

 \thanks{Zhu~Han is with the Department of Electrical and Computer Engineering at the University of Houston, Houston, TX 77004 USA, and also with the Department of Computer Science and Engineering, Kyung Hee University, Seoul, South Korea, 446-701.   (e-mail: hanzhu22@gmail.com).}

 \thanks{Rahim~Tafazolli is with the 5G \& 6G Innovation center (5GIC \& 6GIC), Institute for Communication Systems (ICS), University of Surrey, Guildford, GU2 7XH, U.K., United Kingdom.  (e-mail: r.tafazolli@surrey.ac.uk).}

 \thanks{Tony Q. S.~Quek is with the Singapore University of Technology and Design, Singapore 487372, and also with the Department of Electronic Engineering, Kyung Hee University, Yongin 17104, South Korea (e-mail: tonyquek@sutd.edu.sg).}
 
}

\maketitle

\vspace{-3mm}
\begin{abstract}

The advance of direct satellite-to-device communication has positioned mega-satellite constellations as a cornerstone of 6G wireless communication, enabling seamless global connectivity even in remote and underserved areas. However, spectrum scarcity and capacity constraints imposed by the Shannon's classical information theory remain significant challenges for supporting the massive data demands of multimedia-rich wireless applications. Generative Semantic Communication (GSC), powered by artificial intelligence–based generative foundation models, represents a paradigm shift from transmitting raw data to exchanging semantic meaning. GSC can not only reduce bandwidth consumption, but also enhance key semantic features in multimedia content, thereby offering a promising solution to overcome the limitations of traditional satellite communication systems. This article investigates the integration of GSC into mega-satellite constellations from a networking perspective. We propose a GSC-empowered satellite networking architecture and identify key enabling technologies, focusing on GSC-empowered network modeling and GSC-aware networking strategies. We construct a discrete temporal graph to model semantic encoders and decoders, distinct knowledge bases, and resource variations in mega-satellite networks. Based on this framework, we develop model deployment for semantic encoders and decoders and GSC-compatible routing schemes, and then present performance evaluations. Finally, we outline future research directions for advancing GSC-empowered satellite networks.

	\end{abstract}
	
	\IEEEpeerreviewmaketitle

    \vspace{-3 mm}

	\section{Introduction}
	%
	%
	%
	%

Mega-satellite networks, consisting of thousands of Low Earth Orbit (LEO) satellites, are revolutionizing global connectivity by bridging gaps in underserved regions and delivering seamless network services worldwide. Specifically, according to the International Telecommunication Union (ITU), over 70\% of the Earth's surface, including remote areas such as oceans, deserts, and rainforests, lacks terrestrial network coverage. Furthermore, 2.9 billion people, accounting for 37\% of the global population, still lack Internet access. This significant connectivity gap highlights the critical role of satellite networks in providing reliable and inclusive wireless communication. To fulfill this gap, advances in satellite technology, combined with declining launch costs, are driving the rapid transformation of LEO satellite networks. Among these advancements, key innovations such as inter-satellite links (ISLs) and onboard computing are at the forefront of this evolution.
Compared with traditional Geostationary Earth Orbit (GEO) satellite systems, which incur delays of around 250~ms, LEO satellite networks can achieve much lower latencies in the order of tens of milliseconds by using ISLs to enable data relay between satellites \cite{lai2022spacertc}.
Meanwhile, onboard computing supports local data processing, timely decision-making, and flexible functionality deployment, thereby minimizing reliance on ground stations and enhancing overall network efficiency. Combined with satellite direct-to-device communication capabilities, these advancements aim to support delay-sensitive and bandwidth-intensive applications on a global scale. The increasing demand for services such as real-time Earth observation, Internet of Things (IoT) applications, and satellite-enabled smartphone connectivity further emphasizes the significance of LEO satellite networks. 
As these capabilities evolve, LEO satellite networks are expected to complement ground-based networks, transforming wireless communication across various applications and industries.

While satellite networks promise to support emerging applications, their ability to meet growing demands for high-bandwidth is constrained by limited spectrum and bandwidth resources. Traditional frequency bands, such as L and S, are heavily congested, while C and Ku bands approach saturation. To mitigate these challenges, higher frequency bands like K, Ka  and laser have been adopted \cite{zech2015lct}. Despite these advancements, bandwidth allocation to individual applications remains severely constrained. Techniques like phased array beamforming and dynamic spectrum sharing provide incremental improvements but fail to resolve performance limitations. For instance, tests in March 2024 revealed that download speeds from Starlink satellites to mobile devices were limited to just 17 Mbps, while its initial direct-to-cell constellation offered only 10 Mbps per beam for unmodified cellphones by December 2024. For uplink, the current state-of-the-art achieves only a few Kbps, making it challenging to meet the performance demands of 6G. Furthermore, traditional communication models governed by Shannon's capacity theorem impose fundamental limits on spectrum efficiency, hindering further improvements in wireless communications.

A promising advancement in satellite communication is the integration of generative semantic communication (GSC), which leverages cutting-edge artificial intelligence (AI) to optimize data transmission \cite{deng2024semantic}. Unlike traditional systems that transmit raw data, GSC transmits essential semantic meaning by utilizing a shared knowledge base (KB) between the transmitter and receiver \cite{yang2024rethinking}. This approach ensures that only the most relevant information is conveyed, enabling accurate reconstruction with minimal data \cite{ wang2024harnessing}. Compared to deep learning-based semantic communication, GSC is better suited for mega-satellite networks, as the latter requires distinct models for different channels. Particularly, recent breakthroughs in generative foundation models, such as ChatGPT, have significantly enhanced GSC's capabilities \cite{xu2024unleashing, peng2024semantic}. These models improve semantic understanding, enabling precise reconstruction while reducing data requirements. Their advanced reasoning and context-awareness further allow GSC to effectively manage diverse and dynamic information in wireless communication scenarios \cite{yang2023semantic}. As a result, GSC can offer two key {\em benefits} for satellite networks:

\begin{itemize}

    \item \textbf{Reduced communication load in satellite networks}: By transmitting semantic meaning instead of raw data, GSC can reduce bandwidth usage in satellite networks.

    \item  \textbf{Enhanced  semantic clarity for satellite applications}:   
    By leveraging generative models and semantic reconstruction, GSC enhances the conveyed information. For instance, in satellite-based traffic monitoring systems, traffic signals in raw data may be unclear, but they can become more  recognizable after semantic reconstruction.

\end{itemize}

Additionally, satellite networks can provide unique {\em advantages} for enabling GSC:

\begin{itemize}  

    \item \textbf{Global deployment of GSC:}  
    Satellite networks enable rapid global GSC deployment, ensuring communication in areas lacking ground infrastructure like oceans, deserts, and disaster zones.

    \item \textbf{Minimal resource demands on user devices:}  
    By hosting semantic encoders and decoders on satellites, the need for user devices to handle frequent model deployments or updates is eliminated, reducing computing and communication resource demands. Model updates can be synchronized across satellites, ensuring scalable and resource-efficient GSC system implementation.

\end{itemize}

Despite its benefits, implementing GSC in satellite networks also faces several {\em challenges} from the networking aspect:

\begin{itemize}
\item \textbf{Generative model deployment in satellite networks}: The limited computational resources and dynamic topologies of satellite networks make it challenging to deploy and update generative models efficiently, requiring optimized placement across satellites over time.

\item \textbf{GSC-compatible routing in satellite systems}: 
Traditional routing algorithms primarily focus on the number of nodes and path length. However, integrating GSC into satellite networks presents additional considerations. Semantic encoding and decoding at nodes introduce extra computing latency while reducing bandwidth demand \cite{ren2024generative}. These factors must be thoughtfully addressed in the design of routing algorithms.

\end{itemize}

To tackle these challenges, we propose a GSC-empowered satellite networking architecture, highlighting key enabling technologies, with an emphasis on temporal graph-based network modeling and GSC-compatible networking strategies. We use the discrete temporal graph to capture the diversity of KBs and resource fluctuations in large-scale satellite networks. Building on this architecture, we develop deployment strategies for generative foundation models and GSC-compatible routing schemes, followed by detailed case studies. Finally, we discuss future research directions of GSC-empowered satellite networks for enhancing global wireless connectivity. Our contributions are summarized as follows:

\begin{itemize}
    \item  We propose a GSC-empowered satellite networking architecture and identify key technologies, including temporal graph-based modeling and GSC-compatible networking.
    
    \item  Based on the proposed networking architecture and temporal graph model, we develop deployment strategies for generative foundation models and GSC-compatible routing schemes to enhance network performance.

    \item  We validate the proposed architecture and routing schemes through case studies and outline future directions for advancing GSC-empowered satellite networks.
\end{itemize}

The rest of this article is organized as follows. Section \ref{sec:sys} presents the system architecture and key technologies for the GSC-empowered satellite network. Sections \ref{sec:graph} and \ref{sec:gsc} discuss discrete temporal graph modeling and GSC-compatible model deployment and routing, respectively. A case study on GSC-compatible routing is provided in Section \ref{sec:case}, followed by future challenges and research directions in Section \ref{sec:future}. Finally, the paper concludes with Section \ref{sec:conclusion}.

	
	

    \vspace{-3 mm}

\section{System architecture and key technologies} \label{sec:sys}

\begin{figure}[t!]
	\centering
    \includegraphics[width=3.5in]{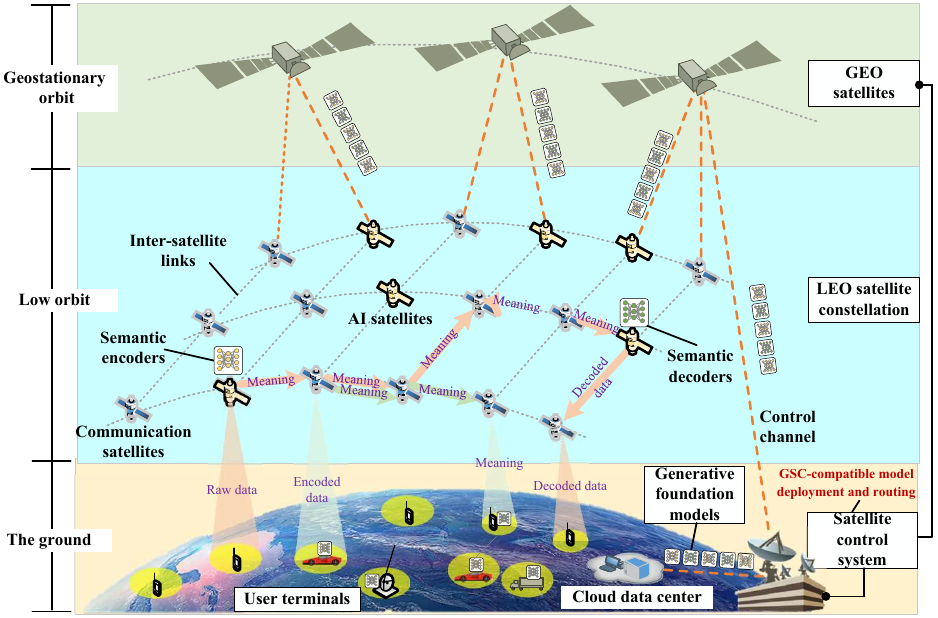}\
	\caption{The overall networking architecture of generative semantic communication (GSC)-empowered satellite networks.}
\label{fig:satellite_semcom_networking_architecture}
        \vspace{-5 mm}
\end{figure}

\subsection{GSC-Empowered Satellite Networking Architecture}

 The GSC-empowered satellite network utilizes GSC to enhance the efficiency of data communications based on generative foundation models. As shown in Fig. \ref{fig:satellite_semcom_networking_architecture}, the overall networking architecture of the GSC-empowered satellite network consists of the LEO satellite constellation, user terminals,  the satellite control system, and generative foundation models serving as the semantic encoders and decoders.

\textbf{LEO Satellite Constellation:} As depicted in Fig. \ref{fig:satellite_semcom_networking_architecture}, the LEO satellite constellation comprises both communication satellites and AI satellites \cite{al2023artificial}. 
The communication satellites are equipped with radio frequency and/or laser transceivers for wireless data forwarding. 
In contrast to communication satellites that focus solely on data transmission, AI satellites have powerful computing capability to support generative foundation model-based semantic encoders and decoders. Specifically, AI satellites leverage resource virtualization technologies to abstract resources into AI functionalities and deploy semantic encoders as microservices, enabling the processing and compression of raw data using semantic encoders corresponding to different KBs. Beyond data transmission, they deploy generative foundation models tailored to specific communication task for encoding and decoding multimedia data, including text, audio, and video. Each AI satellite can function as a  semantic encoder, decoder, or both, depending on computational capacity, power constraints, user demands, and operational costs.
Due to the high costs of satellite launches and designs, the number of AI satellites remains limited, especially in earlier generations \cite{al2023artificial}. Both communication and AI satellites support ISLs, enabling wireless connectivity across the constellation. The network topology evolves predictably due to the periodic nature of satellite orbits. When passing over user terminals, satellites establish satellite-to-ground links (SGLs) to transmit or receive raw, encoded, or decoded data.

\textbf{User Terminals:}
User terminals encompass a wide range of devices, including sensor nodes, vehicle terminals, unmanned aerial vehicles (UAVs), high-altitude platforms, ground stations, and other mobile devices with direct satellite-to-device access. These terminals establish connections with the satellite constellation via SGLs. 
While all these terminals support direct satellite connectivity, their computational capacities vary. Resource-constrained devices like IoT sensors and emergency terminals may lack the power for semantic compression, while others can host full or compressed AI models. However, the frequent model updates required by GSC impose significant computational and communication overhead on user devices. By hosting semantic encoders and decoders on satellites, the burden on resource-limited terminals can be alleviated.

\textbf{Satellite Control System:} 
The satellite control system comprises ground-based satellite control centers (SCCs), GEO satellite systems, and dedicated channels connecting them. The SCCs play a pivotal role in monitoring, managing, deploying, and optimizing the satellite network. They deploy and update semantic encoders and decoders on AI satellites to adapt to generative foundation model improvements, network changes, limited resource constraints, and user demands. To balance computation/communication trade-offs under varying resource conditions, SCCs may selectively deploy models of different sizes based on available computational and communication capabilities. In addition, caching mechanisms can be employed to reduce redundant computations across satellites and over time. Specifically, to handle model updates under latency and power constraints, SCCs first transmit the updated models to GEO satellites, which then relay them to targeted LEO satellites when bandwidth and power availability permit. This GEO-assisted relay mitigates the short contact windows between LEO satellites and ground stations and avoids overloading resource-limited satellites. When onboard resources are constrained, SCCs may reduce the model update frequency or apply model distillation to lower transmission overhead and energy consumption. However, reduced update frequency or model distillation may degrade the accuracy of semantic communication, necessitating careful strategy design to balance resource constraints and network performance.

\textbf{Generative Foundation Models:} 
Generative foundation models are pivotal to the GSC-empowered satellite networking architecture. Deployed on AI satellites and ground terminals, these models are tailored to specific KBs (formally defined in \textbf{Definition~\ref{def:kb}}) to handle diverse data types, such as text, images, and video. Depending on platform resources, they can be deployed in full-scale, fine-tuned, or compressed forms. On AI satellites, these models serve as semantic encoders, decoders, or both, transforming raw data into semantic representations to enhance communication efficiency. Deployment of generative foundation models is resource-dependent, with some satellites hosting multiple models to meet diverse demands. User terminals with sufficient computational capacity can also host generative models for local semantic processing, reducing reliance on satellite resources. To ensure accurate and efficient communication, models are updated and synchronized with the latest KBs. The satellite control system manages updates and deployments, dynamically optimizing resources to adapt to evolving network conditions and user demands.
\begin{definition}[Knowledge Base (KB)]  \label{def:kb}
A KB is a repository of domain-specific knowledge and relevant information that provides shared context between sender and receiver in a semantic communication system. A KB can include text, speech, images, video, or other multimodal datasets, which can be accessed and dynamically updated to support semantic AI model training and inference.
\end{definition}

\textbf{How GSC improves bandwidth efficiency.} GSC reduces bandwidth consumption by transmitting compressed semantic representations instead of raw data, which is particularly beneficial in bandwidth-limited satellite networks. This alleviates traffic on both ISLs and SGLs. However, the dynamic topology of satellite networks requires well-designed networking strategies to fully realize these gains. Without proper coordination, the benefits of GSC may be limited.

 \vspace{-3 mm}

\subsection{Key Technologies of GSC-empowered Satellite Network}

From a networking perspective, the GSC-empowered satellite network involves three key technologies, i.e., GSC-empowered satellite network modeling, generative foundation model deployment, and GSC-compatible routing. Now we will discuss each of them and their related challenges.

\begin{figure}[t!]
	\centering
	  \includegraphics[width=2.8 in]{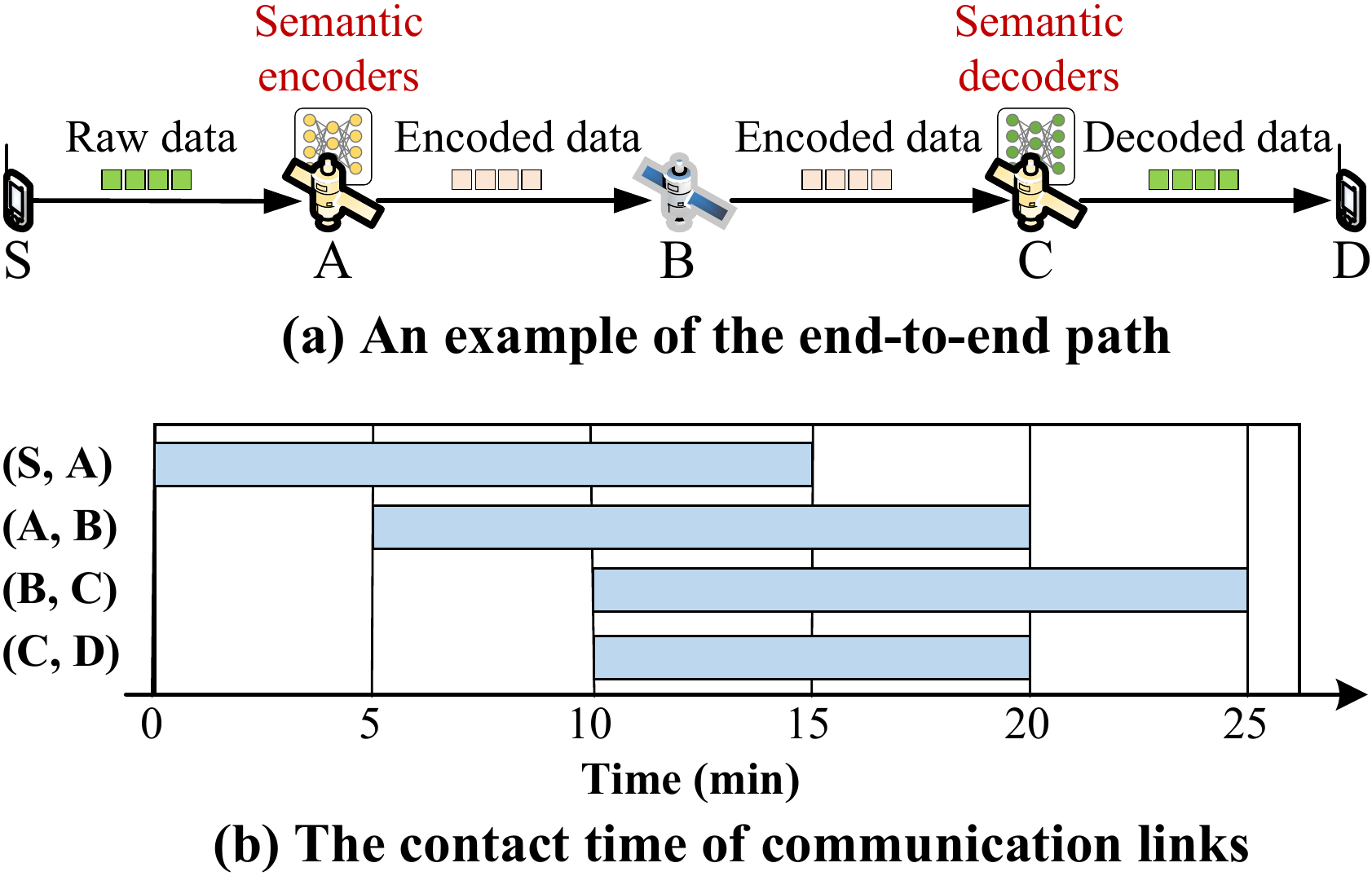}\
	\caption{An example to show the necessity of time discretization in GSC-empowered satellite networks for handling the temporal misalignment of link resources among the end-to-end path. }
	\label{fig:link_aligning_example}
        \vspace{-5 mm}
\end{figure}

\subsubsection{Modeling Time-Varying Node and Link Resources in GSC-empowered Satellite Network} 
Early satellite networks are small in scale and rely on the store-carry-forward model, which was primarily suitable for delay-tolerant applications. In contrast, modern constellations leverage numerous LEO satellites interconnected by ISLs to support real-time communication. However, LEO satellites orbit the Earth every 90 to 120 minutes, maintaining user terminal connectivity for only tens of minutes. This rapid movement causes frequent topology changes and varying link durations, posing significant challenges for network modeling and real-time optimization.

Fig.~\ref{fig:link_aligning_example} presents a data transmission scenario between user terminals \(S\) and \(D\) along the path \( S \to A \to B \to C \to D \) involving three satellites and four different links. The link \( S \to A \) is available from the 0th to the 15th minutes, \( A \to B \) from the 5th to the 20th minutes, \( B \to C \) from the 10th to the 25th minutes, and \( C \to D \) from the 10th to the 20th minutes. Consequently, the path \( S \to A \to B \to C \to D \) remains stable only during the interval from the 5th to the 10th minutes. Traditional static graph models are unable to capture such temporal misalignment among link resources. Therefore, discrete temporal graph models are required to address this limitation by dividing the time axis into discrete time windows, where the network topology within each time window is considered static. This approach ensures alignment of link availability and facilitates efficient resource allocation and routing.

\begin{figure}[t!]
	\centering
	\includegraphics[width=88mm]{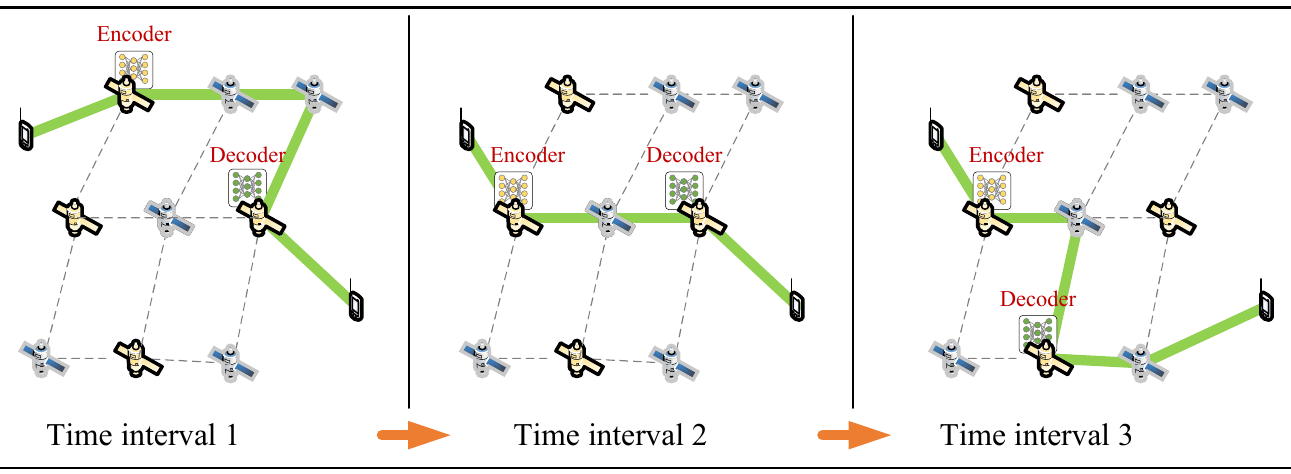}\
	\caption{An example to demonstrate how semantic communication is affected by the movement of satellite nodes. }
	\label{fig:AI_nodes_changes_example}
            \vspace{-5 mm}
\end{figure}

On the other hand, the integration of semantic encoders and decoders into satellite networks necessitates additional optimization of AI satellites while accounting for their mobility. This significantly increases network dynamicity, introducing challenges in efficient resource representation to ensure service continuity. Fig.~\ref{fig:AI_nodes_changes_example} illustrates the impact of satellite mobility on GSC across three intervals. In the first interval, the source user terminal transmits data to an AI satellite for encoding, which relays the encoded data via another satellite to a decoding AI satellite before reaching the destination terminal. In the second interval, the encoding AI satellite moves out of range, prompting the network to reassign both the decoding satellite and relay satellite. In the third interval, further mobility of the decoding satellite necessitates additional updates to sustain communication. The continuous movement of AI satellites introduces substantial complexity in network modeling and resource management, particularly in maintaining continuous semantic communication. Therefore, many critical problems related to network modeling must be addressed:
\begin{itemize}  
    \item How can the dynamic behavior of AI satellites be effectively captured in the network model?  
    \item What strategies can ensure seamless transitions between AI-enabled satellites to avoid communication disruptions?  
    \item How can model deployment schemes be designed to adapt to frequent topological changes and user demands?  
\end{itemize}

\subsubsection{Adaptive Generative Foundation Model Deployment}
Adaptively deploying generative foundation models on satellites is critical for enabling GSC in satellite networks. This involves equipping satellites with semantic encoders and decoders to support diverse user requirements. While certain devices like vehicles can accommodate compressed models using techniques like knowledge distillation, many user terminals lack the computational capabilities for semantic encoding or decoding. For instance, mobile terminals often have limited computational resources, making it challenging to deploy generative foundation models required for tasks like video or 3D volumetric data processing. Additionally, compressed models may compromise accuracy, resulting in ambiguous encoding or imprecise decoding that degrades communication performance. Intuitively, the deployment of GSC components should adhere to fundamental principles to optimize system efficiency. Based on our findings presented in the following lemma, placing the GSC encoder closer to the source node and the decoder nearer to the destination optimizes bandwidth efficiency and minimizes resource consumption. This principle is crucial for designing deployment strategies.

\begin{lemma}[The Fundamental Law of Multi-Hop GSC] \label{lem:fundamental_law}
In a multi-hop GSC-empowered satellite system with a fixed number of hops, positioning the GSC encoder nearer to the source and the GSC decoder nearer to the destination leads to a more efficient utilization of bandwidth during transmission, thereby minimizing the overall resource consumption.
\end{lemma}

\begin{figure}[t!]
	\centering
	 \includegraphics[width=88mm]{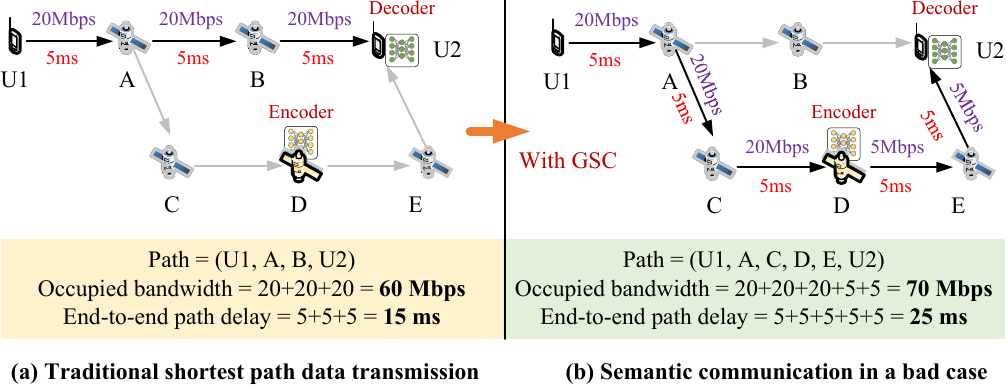}\
	\caption{An example highlighting the importance of network optimization in GSC-empowered satellite networks. 
    }
	\label{fig:bad_case_semantic_communication}
        \vspace{-5 mm}
\end{figure}

Fig.~\ref{fig:bad_case_semantic_communication} presents a concrete example illustrating the importance of strategic deployment. In traditional communication, data is transmitted along the shortest path (U1, A, B, U2), consuming $60$ Mbps bandwidth and incurring a $15$ ms delay. However, if the encoder is deployed on a distant satellite node D, the data must travel through a longer path (U1, A, C, D, E, U2), requiring $70 $ Mbps bandwidth and experiencing a higher delay of $25$ ms. This example underscores how improper deployment can negate the potential benefits of GSC.
As shown in Fig.~\ref{fig:bad_case_semantic_communication}(a), the shortest path latency is 15~ms. Although higher than 5G terrestrial networks, it is much lower than traditional GEO satellites and shows that LEO networks can effectively complement future terrestrial systems in scenarios where terrestrial coverage is absent, such as in remote, underserved, or disaster-affected areas.

Additionally, satellite mobility  presents challenges for GSC model deployment. The periodic movement of satellites can disrupt the optimal placement of encoders and decoders, necessitating adaptive deployment and update strategies.

\subsubsection{GSC-compatible Routing Optimization} 
In GSC systems, routing algorithms are pivotal in managing both data transmission and semantic encoding/decoding processes. However, traditional satellite routing methods are designed with a focus on link-specific metrics like bandwidth and delay. These approaches disregard the semantic requirements intrinsic to GSC, such as where and how data should be encoded or decoded. This limitation renders conventional routing strategies unsuitable for GSC systems.
To address these gaps, routing algorithms tailored to GSC must be developed to ensure efficient and reliable data transmission. The GSC-compatible routing algorithms must resolve several key challenges:  
\begin{itemize}  
    \item \textbf{Matching KBs for Semantic Encoding and Decoding}: 
    The routing algorithm must determine encoding and decoding nodes in addition to selecting the next-hop satellite. For example, if the source node performs semantic encoding, it is crucial to ensure the destination node has compatible decoding capabilities and KBs. Otherwise, satellite-assisted decoding must be utilized.

    \item \textbf{Real-Time Optimization}: To meet the real-time demands of GSC, the algorithm must achieve a balance between bandwidth efficiency and delay, while maintaining low computational complexity for quick responsiveness.

    \item \textbf{Adaptation to Network Dynamics}: Frequent topology changes and rapid mobility of AI nodes require adaptive routing strategies to ensure stable semantic communication under varying conditions.

\end{itemize}  

In conclusion, enabling GSC in satellite networks requires holistic solutions that address network modeling, AI model deployment, and routing optimization. 
These advancements are vital to unlocking the full potential of GSC in dynamic and resource-constrained environments.

\vspace{-3 mm}

\section{Discrete  Temporal Graph Modeling for Dynamic Node and Link Resources in GSC-Empowered Satellite Networks} \label{sec:graph}

Temporal graphs are an effective mathematical tool for capturing the dynamic topological characteristics of satellite networks, particularly in modeling variations in node mobility and communication links. Existing models, such as contact graph \cite{araniti2015contact}, snapshot graph, time-expanded graph, and time-aggregated graph, rely on a time discretization mechanism that divides the time period into discrete intervals  \cite{guo2024enhanced}. During each interval, the network topology (snapshot) is considered to be stable, allowing routing algorithms to function effectively. In existing models, time is divided based on ISL events or fixed time intervals. While they consider satellite mobility and link connectivity, they do not account for the influence of satellite AI capabilities in the context of GSC.

    \vspace{-3 mm}
\subsection{GSC-compatible Time Discretization Method}

In our model, all satellites (both AI satellites and communication satellites) and user terminals are represented as nodes $V$ in the network graph, while ISLs and SGLs form the edges. The KBs deployed in the satellite network are represented by the set $\Theta $, where $|\Theta|$ indicates the total number of supported KBs. For each node $n \in V$, two integer vectors, $\mathbb{E}_n$ and $\mathbb{D}_n$, are defined to characterize its encoding and decoding capabilities with respect to $\Theta$.
A node is classified as a GSC-capable node (either an AI satellite or a user terminal) if at least one element in $\mathbb{E}_n$ or $\mathbb{D}_n$ is  $\geq 1$. Nodes for which all elements of $\mathbb{E}_n$ and $\mathbb{D}_n$ are 0 do not have any deployed encoding or decoding capabilities and function only as communication relays.

Then, the GSC-compatible time discretization process relies on the contact plan \cite{araniti2015contact}, which defines each contact by timestamps (start and end), involved nodes, AI resources, transmission rates, and propagation delays. Due to rapid resource fluctuations in satellite networks, the large number of timestamps often leads to frequent path switching. To address this, we introduce a minimum service duration parameter \( \lambda \), which merges time slots shorter than \( \lambda \) and reconstructs snapshot graphs for the merged intervals. This reduces switching frequency. The process involves two main steps:

\begin{enumerate}

    \item \textbf{Timestamp Sorting}: All contact timestamps are sorted in ascending order to establish initial time intervals.

    \item \textbf{Time Window Discretization}: Time intervals shorter than a predefined threshold \( \lambda \) are merged into larger time windows, reducing frequent state transitions in large-scale satellite networks, as demonstrated in \cite{guo2024enhanced}.

\end{enumerate}

As a result, the time horizon of satellite networks is divided into multiple time windows, where each time window \( \tau \) exceeds the threshold \( \lambda \). By adjusting \( \lambda \),  time windows of varying granularities can be configured for customizing GSC.

\vspace{-3 mm}
\subsection{GSC-compatible Discrete Temporal Graph Construction}

For each time window, a snapshot graph is generated to represent the network state, including only contacts with consistent resources. Each snapshot for time window \( \tau \) is a graph \( {G}^{\tau} = ({V}^{\tau}, {L}^{\tau}) \), where \( {V}^{\tau} \) is the set of  all available nodes, and \( {L}^{\tau} \) includes the available SGLs and ISLs during \( \tau \).

For each AI-capable node in the \( {V}^{\tau}\),the integer indicators of available KBs are stored in a vector, where each element indicates whether a specific KB is available.

\vspace{-3 mm}
\section{GSC-compatible Model Deployment and Routing} \label{sec:gsc}

GSC-compatible model deployment and routing are two critical technologies in GSC-empowered satellite networks. The former serves as a prerequisite for the latter, as efficient model deployment enables optimized routing decisions that adapt to the requirements of user applications. Both GSC-compatible model deployment and routing schemes must be designed in a lightweight manner to reduce computational cost, thereby supporting practical scalability to constellations with thousands of satellites, such as Starlink. 
Furthermore, by optimizing model deployment and routing strategies over the temporal graphs,  these two components collaboratively minimize  bandwidth consumption and traffic load, thereby reducing the overall communication-related energy footprint.

\begin{figure}[!t]
	\centering
    \includegraphics[width=88mm]{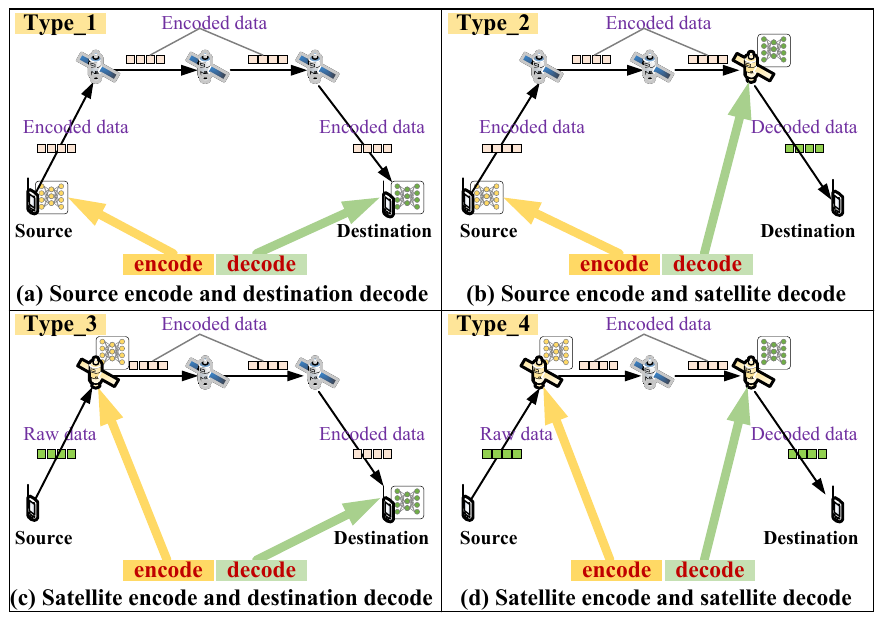}\
    \caption{The four potential types of data communication applications in GSC-empowered satellite networks.
    }
	\label{fig:four_working_modes}
        \vspace{-5 mm}
\end{figure}

    \vspace{-3 mm}

\subsection{Generative Foundation Model Deployment for Semantic Encoders and Decoders}

Realizing GSC in satellite networks requires strategic deployment of generative foundation models to optimize semantic communication, which can be formulated as a mathematical optimization problem. To maximize performance, GSC encoders should be placed near source nodes and decoders near destination nodes. Given predefined transmission requirements, the goal is to meet delay constraints while minimizing bandwidth usage. Binary decision variables can be defined to indicate encoder and decoder deployment on specific satellites, subject to the following constraints:

\begin{itemize}
    \item \textbf{Relay Node Constraints:} Relay nodes must ensure data forwarding integrity by limiting connections, preventing path loops and ensuring continuous data flow.  

    \item \textbf{Link Capacity Constraints:} Communication links must meet the minimum required transmission capacity for GSC, with end-to-end delay within predefined limits.  

    \item \textbf{AI Node Resource Constraints:} Encoder and decoder deployments are constrained by available computational resources on each satellite.  

    \item \textbf{Semantic KB Matching Constraints:} Deployment must ensure that all required KBs are available, ensuring compatibility between source and destination nodes.  
\end{itemize}

Eventually, this optimization problem is modeled as a mixed integer programming problem in a snapshot graph. For multi-snapshot scenarios, custom objectives balance goals like delay minimization and bandwidth efficiency. By solving the instances of the formulated problem across different time windows, we can obtain a dynamic deployment strategy for GSC encoders and decoders, which can be continuously updated. Finally, the SCCs can manage model deployment through control channels, including installing, configuring, and updating encoders and decoders to meet evolving GSC needs.

    \vspace{-3 mm}

\subsection{GSC-compatible Routing for Different Cases}

Efficient routing is essential in supporting GSC, particularly when sender and receiver have different encoding and decoding capabilities. Traditional routing methods fall short in GSC applications, where semantic encoding and decoding must align with the capabilities of nodes.
In Fig.~\ref{fig:four_working_modes}, we classify GSC applications into four types based on different cases  of semantic encoding and decoding capabilities at the user terminals, which require distinct routing strategies as follows:
\begin{itemize}
    \item  \textbf{Case 1: Sender/Receiver Has Encoding/Decoding Capabilities.} This case can be handled by standard shortest-path algorithms under a traditional routing scheme, with no intermediate encoding or decoding involved.

    \item \textbf{Case 2: Sender Encodes, Satellite Decodes.} In this case, the sender encodes the data, and a satellite decodes it for the receiver. The routing algorithm selects the nearest decoding satellite to minimize overall path delay.

    \item \textbf{Case 3: Receiver Decodes, Satellite Encodes.} In this case, since the sender cannot encode the data, the routing algorithm selects the nearest encoding satellite and finds the shortest path to the receiver.

    \item \textbf{Case 4: Both Encoding and Decoding by Satellites.} In this case, encoding and decoding are handled by satellites. The routing algorithm selects one satellite for encoding and another for decoding, while minimizing delay.

\end{itemize}

In summary, each application type requires a tailored routing algorithm that considers the sender, receiver, and satellite capabilities.
According to \textbf{Lemma \ref{lem:fundamental_law}}, when user terminals have sufficient computational resources for running AI models, \textbf{Case 1} is preferred. However, when computing resources on user terminals are limited,  \textbf{Case 2/3/4}  can be selected with satellites handling encoding or/and decoding. Thus, the GSC-compatible routing algorithms must integrate both traditional and GSC methods to maximize network performance while ensuring reliability for different applications in practice.
Moreover, as highlighted in \cite{xu2024unleashing, ren2024generative}, GSC inherently involves computation/communication trade-offs, which also need to be taken into account when designing routing strategies. Larger models yield more accurate semantic representations but incur greater latency and resource costs, while smaller models offer lower overhead at the expense of compromised semantic fidelity. Therefore, an adaptive mechanism must be introduced to adjust model selection based on resource availability, ensuring a balanced trade-off between efficiency and performance in practice.
In addition, caching semantic encoding outputs of frequently accessed content, such as popular videos, at satellites can reduce redundant computation. These outputs can be stored and shared across satellites via ISLs based on predicted user demands to improve efficiency.

\begin{figure*}[!htbp]
\centering
\subfigure[]
  {
  \begin{minipage}{8.6cm}
  \centering
  \includegraphics[width=2.4in]{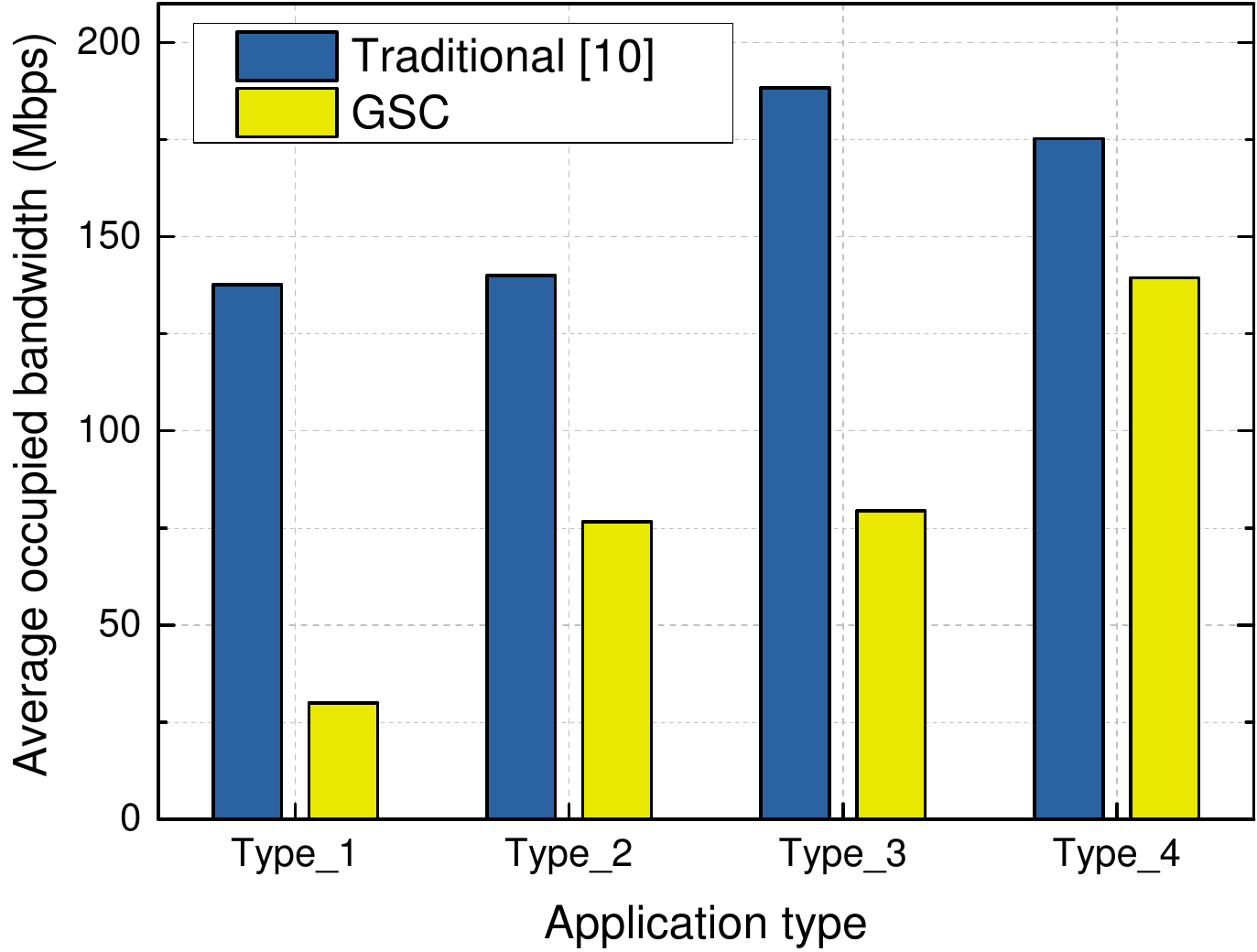}
  \end{minipage}
 }
\subfigure[]
  {
  \begin{minipage}{8.6cm}
  \centering
  \includegraphics[width=2.4in]{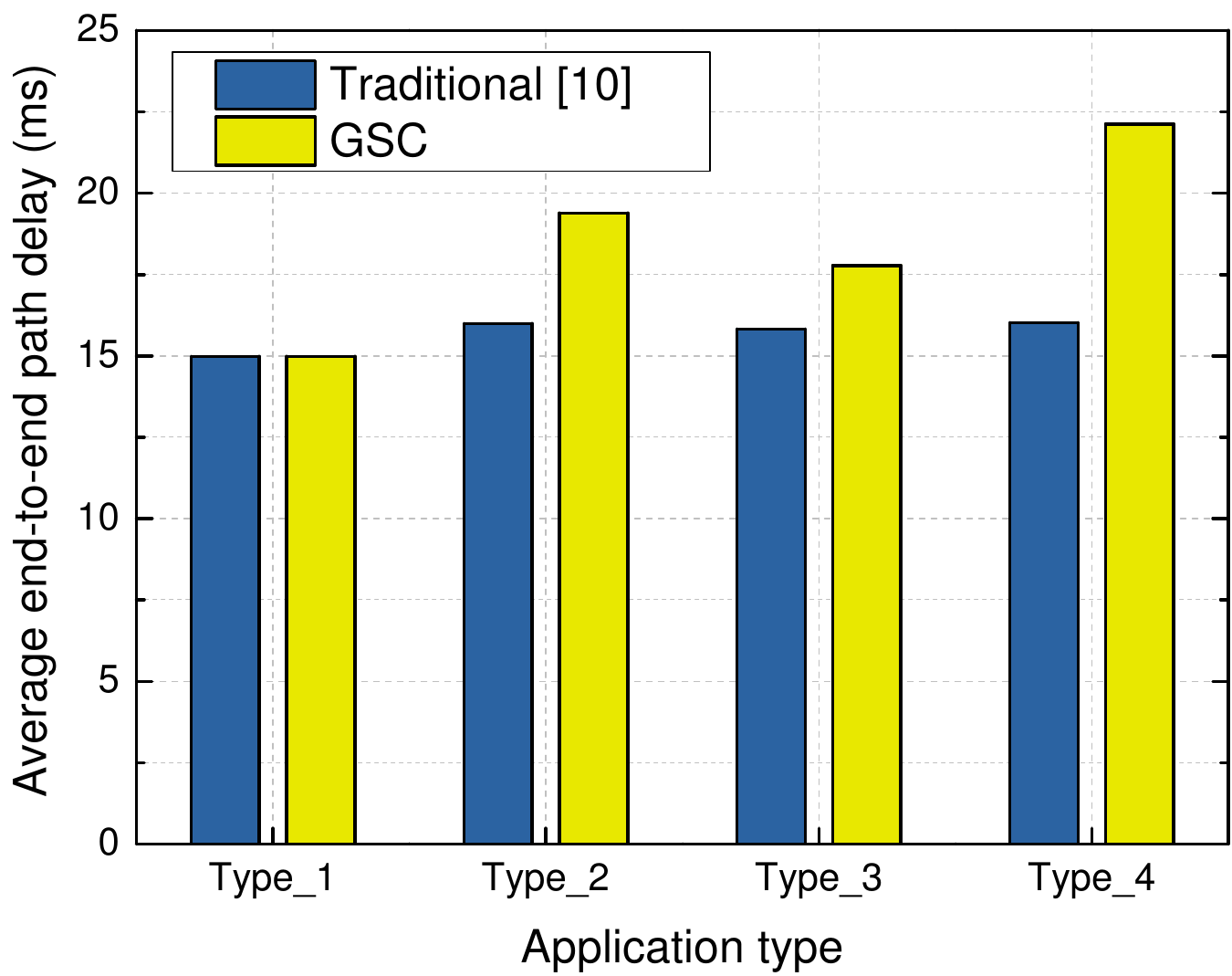}
  \end{minipage}
  }
\caption{a) The average occupied bandwidth of data transmission applications for different routing methods, 
b) The average end-to-end path delay of  data transmission applications for different routing methods.}
\label{fig:6}
        \vspace{-5 mm}
\end{figure*}

    \vspace{-3 mm}

\section{Case Study of GSC-compatible Routing} \label{sec:case}

We evaluate the performance of the GSC-empowered satellite system by simulating 2,500 Starlink satellites and user terminals in 10 regions (Xi'an, Beijing, Sanya, Kashi, Amsterdam, Athens, Barcelona, Berlin, Dubai, and Istanbul). The simulation covers 60 time windows, with 20\% of the Starlink satellites randomly deployed with three different KBs and semantic encoder compression ratios randomly selected from $\{ \frac{1}{8},  \frac{1}{4},  \frac{1}{2}\}$. Four types of applications aforementioned are considered, each with a $25\%$ probability. ISLs and SGLs have bandwidths between 300 and 350 Mbps \cite{fu2020remote} and link delays of 5 to 15 ms \cite{chen2022robust}. We generate 200 applications with random source-destination pairs and data rates from 5 to 100 Mbps.

Fig.~\ref{fig:6}a shows the average occupied bandwidth for four application types under traditional and GSC methods. Occupied bandwidth is calculated as the sum of occupied bandwidth of all used links along the path during transmission. The GSC method reduces bandwidth compared to traditional methods. Type\_1 experiences the largest bandwidth reduction, as both encoding and decoding are performed at the source and destination, respectively, which is the most efficient configuration (as in \textbf{Lemma \ref{lem:fundamental_law}}). For Type\_4, where encoding and decoding occur at the satellite, the reduction is smallest due to the transmission of non-semantic data (raw or decoded) over satellite links. Nevertheless, GSC still achieves a $25\%$ reduction in bandwidth usage across all application types.

Fig.~\ref{fig:6}b shows the average end-to-end path delay for four application types under traditional and GSC methods. Except for Type\_1, where average path delay is equal to the traditional method, average path delays for other types are higher in GSC. Type\_1 can use traditional routing without encoding or decoding at the satellite, while the need to locate AI nodes for the other types leads to longer paths and increased delays. This shows that GSC reduces bandwidth but increases delay as a trade-off. 
For type\_4, due to limited computing capabilities of user devices, both semantic encoding and decoding must be performed on satellites. As a result, raw data must be transmitted over the SGLs, while only the ISLs benefit from semantic compression. In this case, deploying GSC models on satellites closer to user devices can further enhance performance.

\vspace{-3 mm}

\section{Future Challenges and Research Directions} \label{sec:future}

The emergence of GSC represents a major shift in satellite communication. By hosting semantic encoders and decoders, satellite networks can enable global GSC implementation, reducing the burden on resource-constrained devices and ensuring bandwidth-efficient communication. However, several key challenges must be addressed to fully realize GSC's potential in satellite networks.

\textbf{Standardization of Networking Framework.}
For GSC' s widespread adoption in satellite networks, a standardized networking framework is crucial. This framework should define common benchmarks, performance metrics, and interfaces to ensure interoperability across satellite networks and user terminals. By establishing network protocols for semantic encoding, decoding, and model updates, standardization will streamline GSC integration, improve scalability, and foster collaboration among satellite operators and service providers.

\textbf{Testbed Evaluation of Networking Strategies.}
Establishing testbeds for evaluating GSC in satellite networks is a significant challenge, primarily due to the distributed and dynamic nature of satellite networks. While network virtualization offers flexibility in simulating different scenarios \cite{zheng2023sdn}, further research is necessary to develop emulators for testing GSC in real-world conditions. In particular, optimizing applications for satellite environments and tailoring model deployment strategies are crucial for ensuring the successful integration of GSC into satellite networks. Such testbeds will not only allow for the validation of GSC techniques but also help identify optimal networking strategies for managing resources and improving network performance.

\textbf{Security and Resilience of GSC-empowered Satellite Systems.}
Integrating GSC into satellite systems introduces complexities such as data unpredictability and manipulation risks. Given the reliance on real-time model updates, safeguarding data integrity is crucial. Strong security measures, including error detection and privacy-preserving techniques, are needed to ensure reliable and secure data transmission. Additionally, resilience strategies are essential to protect against network disruptions and attacks, ensuring continuous performance.

\vspace{-3mm}
\section{Conclusion} \label{sec:conclusion}

This article has explored the integration of GSC into mega-satellite constellations from a networking perspective. We have proposed a GSC-empowered satellite networking architecture and identify enabling technologies, focusing on GSC-based network modeling and routing strategies. A discrete temporal graph has been introduced to capture the diversity of KBs and the dynamics of communication resources in these networks. Using this framework, we have developed and evaluated a GSC-compatible routing scheme. Finally, we have highlighted future research directions to advance GSC in satellite networks.
Future work will elaborate on lightweight GSC-compatible model deployment and routing methods to enhance scalability and reduce energy footprint, while also balancing computation/communication trade-offs and leveraging caching to minimize redundant computation in large satellite constellations.

\vspace{-3mm}

\section{Acknowledgement}
This work is supported by the National Natural Science Foundation of China under Grant 62171456, in part by the National Research Foundation, Singapore and Infocomm Media Development Authority under its Future Communications Research $\&$ Development Programme, in part by the SNS JU project 6G-GOALS under the EU' s Horizon program Grant Agreement No. 101139232, in part by NSF ECCS-2302469, CMMI-2222810, Toyota. Amazon and Japan Science and Technology Agency (JST) Adopting Sustainable Partnerships for Innovative Research Ecosystem (ASPIRE) JPMJAP2326, and in part by the Fund of China Scholarship Council.

\bibliographystyle{IEEEtran}
\bibliography{reference}
\vspace{-12 mm}
\begin{IEEEbiographynophoto}
\indent \textbf{Binquan Guo} (bqguo@stu.xidian.edu.cn) received the B.S. and M.S. degree in telecommunication engineering from Xidian University in 2017 and 2020, respectively. He is currently pursuing the Ph.D. degree with the State Key Laboratory of Integrated Service Networks, Xidian University, and the Tianjin Artificial Intelligence Innovation Center. He is a visiting student at SUTD, Singapore, sponsored by the CSC. His research interests include satellite networking, network slicing, graph theory, routing and scheduling. \\

\indent \textbf{Wanting Yang}  (wanting\_yang@sutd.edu.sg) received the B.S. and Ph.D. degrees from the Department of Communications Engineering, Jilin University, Changchun, China, in 2018 and 2023, respectively. She is currently a Research Fellow with the Singapore University of Technology and Design. Her research interests include wireless semantic communication, generative artificial intelligence, learning, martingale, and predictive resource allocation. 

~

\indent \textbf{Zehui Xiong} (z.xiong@qub.ac.uk) (Senior Member, IEEE) is currently a full professor with the School of Electronics, Electrical Engineering and Computer Science, Queen's University Belfast. His research interests include wireless communications, network games and economics, blockchain, and edge intelligence.

~

\indent \textbf{Zhou Zhang} (zt.sy1986@163.com) (Member, IEEE) received the Ph.D. degree in electrical engineering from the University of Alberta, Edmonton, AB, Canada, in 2013. He is currently with College of Computer Science and Electronic Engineering, Hunan University, Changsha, P. R. China. His research interests include cooperative communications and radio resource management for 5G wireless networks and beyond.

~

\indent \textbf{Baosheng Li } (bs.li@stu.xidian.edu.cn) is currently pursuing the Ph.D. degree with the School of Mathematics and Statistics, Xidian University, China. His research interests include distributed learning and federated optimization.

~

\indent \textbf{Zhu Han} (Fellow, IEEE) (hanzhu22@gmail.com) received his Ph.D. from the University of Maryland. He is a John and Rebecca Moores Professor in the Electrical and Computer Engineering Department and Computer Science Department at the University of Houston, Texas. He has been an AAAS Fellow since 2019, and an ACM Fellow since 2024.  He is the winner of 2021 IEEE Kiyo Tomiyasu Award (IEEE Technical Field Awards).

~

\indent \textbf{Rahim Tafazolli} (Fellow, IEEE), CBE,  Regius Professor of Electronic Engineering, (r.tafazolli@surrey.ac.uk) is a Fellow of Royal Academy of Engineering (FREng) , FIET, FCIC, Fellow of WWRF and Professor of Mobile and Satellite Communications, Founder and Director of 5GIC, 6GIC and ICS (Institute for Communication System) at the University of Surrey. He has over 30 years of experience in digital communications research and teaching. He has authored and coauthored more than 1000 research publications and is regularly invited to deliver keynote talks and distinguished lectures to international conferences and workshops.

~

\indent \textbf{Tony Q.S. Quek}(S'98-M'08-SM'12-F'18) (tonyquek@sutd.edu.sg) received the B.E.\ and M.E.\ degrees in electrical and electronics engineering from the Tokyo Institute of Technology in 1998 and 2000, respectively, and the Ph.D.\ degree in electrical engineering and computer science from the Massachusetts Institute of Technology in 2008. Currently, he is the Associate Provost (AI \& Digital Innovation) and Cheng Tsang Man Chair Professor with Singapore University of Technology and Design (SUTD). He also serves as the Director of the Future Communications R\&D Programme, and the ST Engineering Distinguished Professor. He was honored with the 2008 Philip Yeo Prize for Outstanding Achievement in Research, the 2012 IEEE William R. Bennett Prize, the 2017 CTTC Early Achievement Award, the 2017 IEEE ComSoc AP Outstanding Paper Award, the 2020 IEEE Communications Society Young Author Best Paper Award, the 2020 IEEE Stephen O. Rice Prize, the 2022 IEEE Signal Processing Society Best Paper Award, the 2024 IIT Bombay International Award For Excellence in Research in Engineering and Technology, and the IEEE Communications Society WTC Recognition Award 2024. He is an IEEE Fellow, a WWRF Fellow, and a Fellow of the Academy of Engineering Singapore.

\end{IEEEbiographynophoto}

\end{document}